\documentclass[prb,superscriptaddress,twocolumn,
longbibliography]{revtex4-2}

\usepackage{amsmath,amssymb}
\usepackage{physics}
\usepackage{graphicx}
\usepackage{bm}
\usepackage{bbm}
\usepackage[colorlinks,bookmarks=false,citecolor=blue,linkcolor=black,urlcolor=red]{hyperref}
\usepackage{color} 
\usepackage{newtxtext,newtxmath}

\allowdisplaybreaks

\begin{document}

\title{
Subsystem many-hypercube codes: High-rate concatenated codes with low-weight syndrome measurements
}

\date{\today}

\author{Ryota Nakai}
\affiliation{RIKEN Center for Quantum Computing (RQC), Wako, Saitama, 351-0198, Japan}

\author{Hayato Goto}
\affiliation{RIKEN Center for Quantum Computing (RQC), Wako, Saitama, 351-0198, Japan}
\affiliation{Corporate Laboratory, Toshiba Corporation, Kawasaki, Kanagawa 212-8582, Japan}

\begin{abstract}
 Quantum error-correcting codes (QECCs) require high encoding rate in addition to high threshold unless a sufficiently large number of physical qubits are available.
 The many-hypercube (MHC) codes defined as the concatenation of the $[\![6,4,2]\!]$ quantum error-detecting code have been proposed as high-performance and high-encoding-rate QECCs.
 However, the concatenated codes have a disadvantage that the syndrome weight grows exponentially with respect to the concatenation level.
 To address this issue, here we propose subsystem quantum codes based on the MHC codes. 
 In particular, we study the smallest subsystem MHC codes, namely, subsystem codes derived from the concatenated $[\![4,2,2]\!]$ error-detecting codes.
 The resulting codes have a constant syndrome-measurement weight of 4, while keeping high encoding rates.
 We build the block-MAP and neural-network decoders and show that they demonstrate superior performance to the bounded-distance decoder.
\end{abstract}

\maketitle

%\tableofcontents

\section{Introduction}

Quantum error-correcting codes (QECCs) \cite{GottesmanWeb,Nielsen_Chuang_2010} are widely accepted as a means to perform quantum computations fault-tolerantly.
A quantum circuit on physical qubits are encoded into that on logical qubits,
which are protected from noise by the repetition of detecting and correcting errors via syndrome measurements.
The surface code is the current standard QECC \cite{KITAEV20032,bravyi1998quantumcodeslatticeboundary,PhysRevA.86.032324}.
The advantage of the surface code lies in the fact that it has a high error threshold and is implemented locally via nearest-neighbor interaction on a two-dimensional array of qubits. 
However, fault-tolerant quantum computations with it require high space overhead, that is, one needs $d^2$ physical qubits to encode a single logical qubit with code distance $d$. 
The bivariate bicycle (BB) codes, which are representative of the quantum low-density parity-check (qLDPC) codes, have been proposed as a solution to this problem \cite{Bravyi2024}.
With the same code distance, the encoding rate is much higher than that of the surface code. 
The syndrome measurements in the BB code involve weight-6 check operators, but, unlike the surface code, they are non-local: one needs to interact two qubits far apart.

Another solution to the encoding-rate problem has been proposed based on code concatenation \cite{Yamasaki2024,doi:10.1126/sciadv.adp6388,Yoshida2025}.
Here, we refer to the number of codes that are concatenated to construct a single code block as concatenation level or simply level, which is denoted by $r$ in the following.
The concatenation of the quantum Hamming codes \cite{PhysRevA.54.1098,PhysRevA.54.4741} of sizes $7, 15, 31,\cdots$ up to $2^{r+2}-1$ leads to an asymptotically constant encoding rate of 1/36 as the concatenation level increases \cite{Yamasaki2024}.
The concatenation of the $[\![6,4,2]\!]$ quantum error-detecting code, the resultant codes of which are referred to as the $[\![6,4,2]\!]$ many-hypercube (MHC) codes \cite{doi:10.1126/sciadv.adp6388,liu2025coniqenablingconcatenatedquantum,rines2025demonstrationlogicalarchitectureuniting}, has also been proposed.
Within intermediate sizes, the $[\![6,4,2]\!]$ MHC codes have high encoding rate of $(4/6)^{r}$.
In such concatenated codes, however, the weights of the syndromes (check operators) grow exponentially as the concatenation level increases.
This is because the weights of lower-level logical operators, of which the check operators consist, also grow exponentially.
Specifically, the level-$r$ concatenated quantum Hamming code has check operators of weight as high as $4\cdot 6^{r-1}$ \cite{Yamasaki2024}, and the level-$r$ $[\![6,4,2]\!]$ MHC code has those of weight $6\cdot 2^{r-1}$ \cite{doi:10.1126/sciadv.adp6388}.

A similar problem occurred in the Shor code \cite{PhysRevA.52.R2493}, which is a concatenated code of the 3-qubit repetition codes for bit-flip and phase-flip errors.
When the former code is defined using logical qubits of the latter code, the syndrome measurements for bit-flip errors are performed on 6 qubits, while those for phase-flip errors on 2 qubits.
The weight of the check operators can be reduced by making part of the qubits not engaged in quantum computations.
The resulting code is known as the Bacon-Shor code, which is a type of the subsystem code \cite{PhysRevLett.84.2525,PhysRevLett.95.230504,PhysRevA.73.012340}.
The redundant part of the qubits are referred to as the gauge qubits.
Since any two operators are equivalent modulo gauge-qubit operators, the weight of the check operators is reduced to 2 in the Bacon-Shor code.
There is another method of reducing the check-operator weight by introducing time periodic syndrome measurements, known as the Floquet codes 
\cite{Hastings2021dynamically,Haah2022boundarieshoneycomb,Gidney2021faulttolerant,Gidney2022benchmarkingplanar,EPTCS384.14,Bombin2024unifyingflavorsof,PRXQuantum.4.020341,PRXQuantum.5.020305,alam2024dynamicallogicalqubitsbaconshor}.

In this work, we propose subsystem codes made from the MHC codes \cite{doi:10.1126/sciadv.adp6388}, which are referred to as the subsystem MHC codes.
As a simple example, we study the $[\![4,2,2]\!]$ subsystem MHC codes, which are the subsystem codes of the concatenated $[\![4,2,2]\!]$ error-detecting codes.
The $[\![4,2,2]\!]$ code is the smallest stabilizer code that can detect a single-qubit error \cite{PhysRevA.54.R1745,PhysRevA.56.33}.
The resulting code at concatenation level $r$ is a $[\![4^r,2^r,g,2^r]\!]$ code, where $g=4^r+2^r-2\cdot3^r$ is the number of gauge qubits.
The syndrome measurements are performed on four qubits at an arbitrary level,
while the original level-$r$ $[\![4,2,2]\!]$ MHC code has check operators of weight as high as $4\cdot 2^{r-1}$.
However, the decoding becomes problematic due to a large number of gauge qubits.
We overcome this problem by building two kinds of decoders for the new codes: a block MAP (maximum-a-posteriori) decoder \cite{PhysRevA.74.052333} up to level 3 and a neural-network (NN) decoder \cite{PhysRevA.99.052351,Baireuther_2019} up to level 4. 
We evaluate the performance of the $[\![4,2,2]\!]$ subsystem MHC codes and show that they outperform the theoretical performance of the bounded-distance (BD) decoder and also the performance evaluated with the belief propagation with ordered statistics decoding (BP-OSD) \cite{Panteleev2021degeneratequantum}.

In the following, we denote the $[\![4,2,2]\!]$ and $[\![6,4,2]\!]$ error-detecting codes by $D_4$ and $D_6$, respectively,
and their concatenated codes by $D_{n_1,n_2,\cdots,n_r}$ when the level-$i$ code is $D_{n_i}$ following the notation of \cite{liu2025coniqenablingconcatenatedquantum}.
Specifically, the original level-$r$ $[\![4,2,2]\!]$ MHC codes are denoted by $D_{4^r}$ (e.g. $D_{4,4}$ for level 2 and $D_{4,4,4}$ for level 3)
and we refer to the level-$r$ subsystem ones as the subsystem $D_{4^r}$ codes.

The rest of this paper is organized as follows.
In Sec.~\ref{sec:code}, we explain the definition of the subsystem $D_{4^r}$ codes in more detail.
In Sec.~\ref{sec:decoder}, we present our decoders and show their performance.
Finally, we give conclusions in Sec.~\ref{sec:conclusion}.

\section{[[4,2,2]] subsystem MHC codes}
\label{sec:code}

\subsection{$D_{4}$ code}

\begin{figure}
    \centering
    \includegraphics[width=0.36\textwidth]{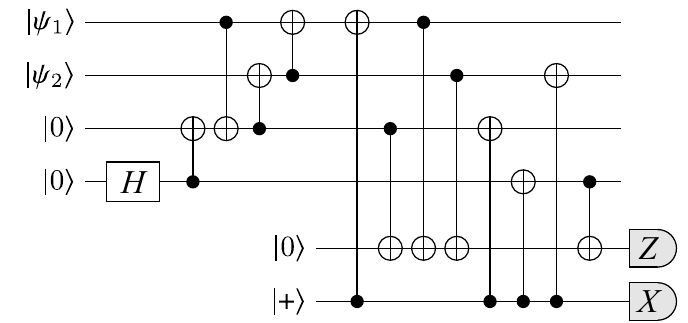}
    \caption{
        A circuit for an encoder of $|\psi_1\psi_2\rangle_L$ state and the syndrome measurement of the $D_4$ code.
        \label{fig:422level1}
    }
\end{figure}

We start with the definition of the $D_4$ code \cite{PhysRevA.54.R1745,PhysRevA.56.33}.
The $D_4$ code is a stabilizer code defined on four qubits
and has two stabilizer generators given by
\begin{align}
    SX^{(1)}=X_1X_2X_3X_4,\quad SZ^{(1)}=Z_1Z_2Z_3Z_4,
    \label{eq:stabilizers_level1}
\end{align}
where $X_j$ and $Z_j$ are the Pauli operators on qubit $j$, and the number in the superscript indicates the level of concatenation.
The logical operators of the encoded two qubits are given by
\begin{align}
    \left\{
    \begin{array}{l}
        X^{(1)}_{L1}=X_2X_3\\
        Z^{(1)}_{L1}=Z_1Z_2
    \end{array}
    \right.,\quad
    \left\{
    \begin{array}{l}
        X^{(1)}_{L2}=X_1X_2\\
        Z^{(1)}_{L2}=Z_2Z_3
    \end{array}
    \right..
    \label{eq:logicals_level1}
\end{align}
The code space is spanned by
\begin{align}
    |\psi_1\psi_2\rangle_L
    =
    \left(X^{(1)}_{L1}\right)^{\psi_1}
    \left(X^{(1)}_{L2}\right)^{\psi_2}|00\rangle_L,
\end{align}
where $\psi_1,\psi_2 = 0,1$ and 
$
    |00\rangle_L
    =
    (|0000\rangle+|1111\rangle)/\sqrt{2}
$
is the GHZ state.
A quantum circuit for an encoder \cite{Knill2005} followed by the syndrome measurement  of (\ref{eq:stabilizers_level1}) using two ancilla qubits \cite{Reichardt_2021} for the $D_4$ code is shown in Fig.~\ref{fig:422level1}.
This syndrome extraction method can detect harmful errors that spread into data qubits \cite{Reichardt_2021}.

\subsection{$D_{4,4}$ code}

Let the four physical qubits of the $D_4$ code be aligned along the $x$ axis from $x=1$ to $4$.
At level 2, we prepare four sets of the $D_4$ code, each of which is specified by the position along the $y$ axis from $y=1$ to $4$.
The $D_{4,4}$ code is then defined on $4^2=16$ physical qubits on two-dimensional lattice sites \cite{doi:10.1126/sciadv.adp6388}.

\begin{figure*}
    \centering
    \includegraphics[width=1\textwidth]{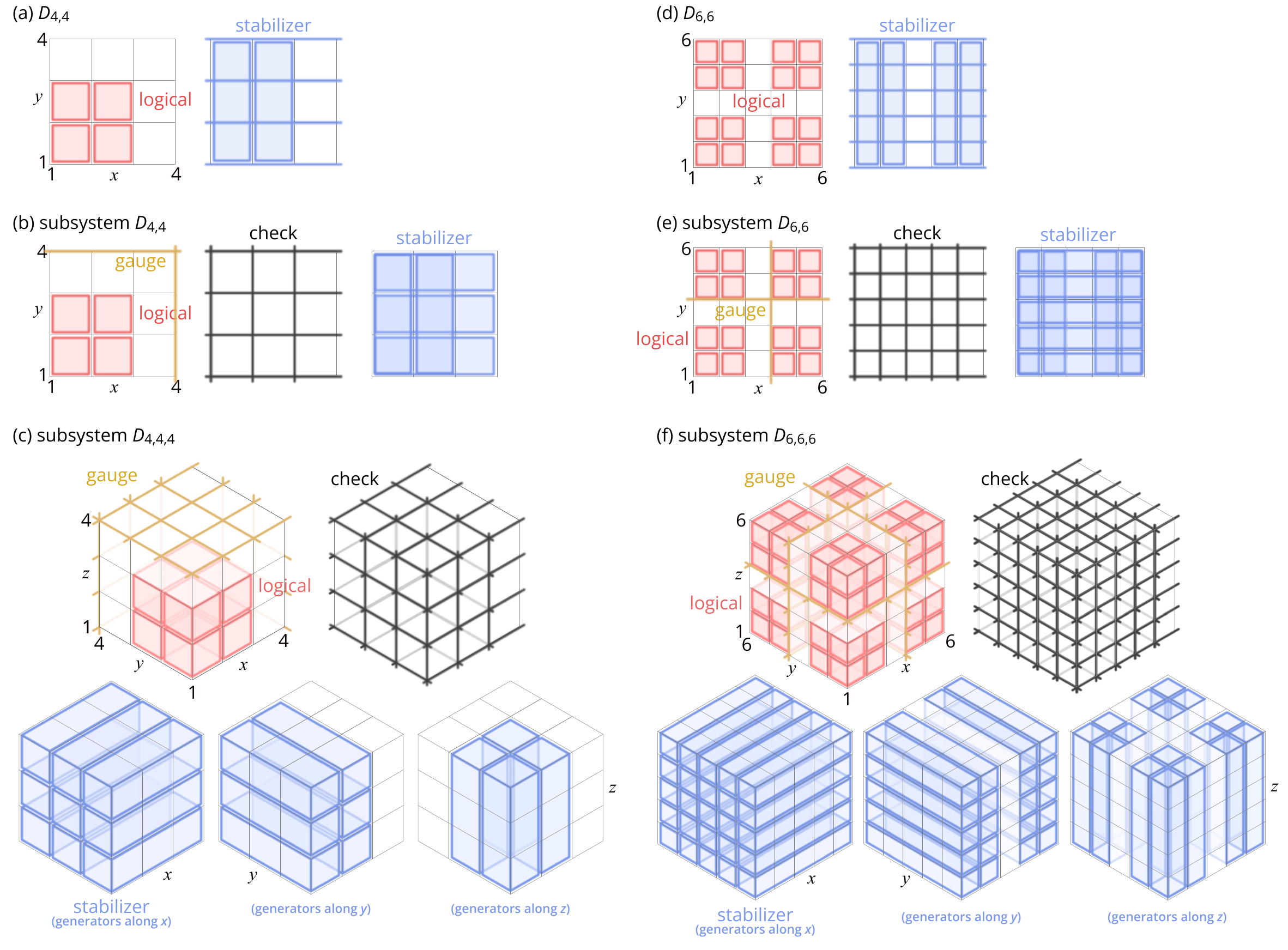}
    \caption{
        The logical-qubit operators, gauge-qubit operators, stabilizer generators and check operators of (a) the $D_{4,4}$, %code and 
        (b) subsystem $D_{4,4}$, 
        (c) subsystem $D_{4,4,4}$,
        (d) $D_{6,6}$,
        (e) subsystem $D_{6,6}$, and
        (f) subsystem $D_{6,6,6}$ codes are shown.
        In (c) and (f), the stabilizer generators that extend along the $x$, $y$, and $z$ directions are shown separately.
        The logical-qubit (gauge-qubit) operators are defined on qubits shown by red squares or cubes (yellow lines).
        Notice that the logical $X$ and $Z$ operators of the same logical or gauge qubits are not defined on the same sets of qubits; they share an odd number of qubits so that they anticommute with each other.
        Two types of stabilizer generators $SX$ and $SZ$ are defined on the same sets of qubits shown by blue lines, rectangles or cuboids.
        This is also the case for the $X$ and $Z$ check operators (black lines).
        \label{fig:cube}
    }
\end{figure*}

The logical operators and part of the stabilizer generators of $D_{4,4}$ are constructed using the logical operators of $D_4$.
There are two logical qubits per each $D_4$ code, for a total of 8 logical qubits.
The $D_{4,4}$ code consists of two $D_4$ codes, each of which is defined by four level-1 logical qubits.
Let the $j$th level-1 logical operators at $y$ be denoted by $X_{Lj,y}^{(1)}$ and $Z_{Lj,y}^{(1)}$, where $j\in [1,2]$ and $y\in[1,4]$.
One obtains the level-2 stabilizer generators [Fig.~\ref{fig:cube} (a)]
\begin{align}
    SX_j^{(2)y}=\bigotimes_{y=1}^4X_{L3-j,y}^{(1)},\quad
    SZ_j^{(2)y}=\bigotimes_{y=1}^4Z_{Lj,y}^{(1)},
    \label{eq:stabilizers_manyhypercubelevel2}
\end{align}
and the logical operators
\begin{align}
    X^{(2)}_{Ljk}=\bigotimes_{y=3-k}^{4-k}X^{(1)}_{Lj,y},\quad 
    Z^{(2)}_{Ljk}=\bigotimes_{y=k}^{k+1}Z^{(1)}_{Lj,y},
    \label{eq:level2_logicals}
\end{align}
where $j,k=1,2$.
Here $y$ in the superscript of (\ref{eq:stabilizers_manyhypercubelevel2}) indicates that the stabilizer generators extends along the $y$ axis [Fig.~\ref{fig:cube} (a)].
The stabilizer generators with the same subscript are made to be defined on the same set of physical qubits.
Combined with the level-1 stabilizer generators, there are 12 stabilizer generators and 4 logical qubits that have weight-4 logical operators, forming a $[\![4^2,2^2,2^2]\!]$ code.
8 out of 12 stabilizer generators are weight-4 operators and the remaining ones given by (\ref{eq:stabilizers_manyhypercubelevel2}) are weight-8 operators [Fig.~\ref{fig:cube} (a)].

\subsection{Subsystem $D_{4,4}$ code}

A $[\![n,k,g,d]\!]$ subsystem code encodes $k$-qubit states on $n$ physical qubits with code distance $d$.
It contains $g$ gauge qubits. 
We construct a $[\![4^2,2^2,2,2^2]\!]$ subsystem code from the $D_{4,4}$ code by
assigning four-qubit operators at $x=4$ and $y=4$ as the gauge-qubit operators [Fig.\ref{fig:cube} (b)].
Specifically, they are given by the level-1 stabilizer generators $SX_4^{(1)},SZ_4^{(1)}$ at $y=4$ and two operators
\begin{align}
    \bigotimes_{y=1}^4X_{4y},\,\,\bigotimes_{y=1}^4Z_{4y}
\end{align}
at $x=4$, where $X_{xy}$ and $Z_{xy}$ are Pauli operators of the physical qubit at $(x,y)$.
The gauge-qubit operators do not share physical qubits with the level-2 logical-qubit operators (\ref{eq:level2_logicals}).
Since these operators are now irrelevant to the code, the level-2 stabilizer generators (\ref{eq:stabilizers_manyhypercubelevel2}) are equivalent to weight-four check operators [Fig.~\ref{fig:cube} (b)].
As a result, there are 14 independent check operators with weight four in the subsystem $D_{4,4}$ code.

The check operators form a gauge group $\mathcal{G}$, from which a subsystem code is defined \cite{PhysRevLett.95.230504,alam2024dynamicallogicalqubitsbaconshor}.
The stabilizer group is the Abelian subgroup of $\mathcal{G}$ defined by $\mathcal{Z}(\mathcal{G})\cap\mathcal{G}$, where $\mathcal{Z}(\mathcal{G})$ is the centralizer of $\mathcal{G}$ \cite{Nielsen_Chuang_2010}.
The stabilizer generators are weight-8 operators, which are the products of two adjacent parallel check operators [Fig.~\ref{fig:cube} (b)].
Therefore, we can determine the syndrome outcome by measuring the weight-four check operators and calculating the products of the measurement outcome.
Notice that while the measurement outcome of the check operators is determined probabilistically since they are not commutable with each other, their products are not.
There are 10 stabilizer generators in the subsystem $D_{4,4}$ code.
The logical-qubit operators are part of the generators of $\mathcal{Z}(\mathcal{G})$ that are not included in $\mathcal{G}$, and are the same as those of the $D_{4,4}$ code (\ref{eq:level2_logicals}).

\subsection{Subsystem $D_{4^r}$ codes}
\label{sec:levelr}

The subsystem $D_{4^r}$ code is defined on $4^r$ qubits forming an $r$-dimensional hyperlattice of linear dimension three.
Along each axis, there are $4^{r-1}$ parallel lines that connect four qubits, for a total of $r\cdot 4^{r-1}$ lines.
We define weight-four check operators as the products of $X$ and $Z$ on these lines, respectively.
The gauge group $\mathcal{G}$ is generated by the check operators.
The stabilizer generators of $\mathcal{S}=\mathcal{Z}(\mathcal{G})\cap\mathcal{G}$ are operators on hypercuboids of $4\cdot 2^{r-1}$ physical qubits (4 in one of the $r$ directions and 2 in the other ones).
For each direction, there are $2\cdot 3^{r-1}$ stabilizer generators.
However, not all the stabilizer generators defined in this way are independent; there are $2(3^r-2^r)$ independent stabilizer generators in the subsystem $D_{4^r}$ code.
The syndrome outcome of the stabilizer generators (weight $2^{r+1}$) is determined by the products of the measurement outcome of $2^{r-1}$ check operators (weight four).
This indicates that the measurement weight is reduced from $2^{r+1}$ to four by making the $D_{4^r}$ code into a subsystem code.

The remaining generators of $\mathcal{Z}(\mathcal{G})$ not included in $\mathcal{S}$ forms $2^r$ logical qubits, whose operator weight is $2^r$.
The remaining generators of $\mathcal{G}$ not included in $\mathcal{S}$ forms gauge qubits.
In total, there are $2^r$ logical qubits, $2(3^r-2^r)$ stabilizer generators, and $4^r+2^r-2\cdot3^r$ gauge qubits.

The operators of the subsystem $D_{4^r}$ code is defined recursively from the lower level codes.
Let the stabilizer generators and logical-qubit operators of the subsystem $D_{4^{r-1}}$ code at $x_r\in[1,4]$ be denoted by
\begin{align}
    SA_{j_1\cdots \check{j}_n\cdots j_{r-1},x_r}^{(r-1)x_n},\,\, 
    A_{Lj_1\cdots j_{r-1},x_r}^{(r-1)},
\end{align}
where $n\in[1,r-1]$ is the direction along which the operator extends
and $A=X,Z$.
Here, $j_1\cdots \check{j}_n\cdots j_r$ indicates a set of indices obtained from $j_1\cdots j_r$ by omitting $j_n$,
where $j_1,\cdots, j_{n-1}$ take 1 and 2 while $j_{n+1},\cdots, j_{r}$ run over 1 to 3.
On the other hand, the subscript indices of the logical operators take 1 and 2.

From these operators, the stabilizer generators at level $r$ are given by
\begin{align}
    &SA_{j_1\cdots \check{j}_n\cdots j_r}^{(r)x_n}=\bigotimes_{x_r=j_r}^{j_r+1} SA_{j_1\cdots \check{j}_n\cdots j_{r-1},x_r}^{(r-1)x_n} \label{eq:stabilizerr_from_stabilizerr-1}
\end{align}
for $n\in[1,r-1]$, $j_r\in[1,3]$ and 
\begin{align}
    &SX_{j_1\cdots j_{r-1}\check{j}_r}^{(r)x_r}=\bigotimes_{x_r=1}^{4}X_{L3-j_1\cdots 3-j_{r-1},x_r}^{(r-1)}, \label{eq:stabilizerr_from_logicalr-1_0}\\
    &SZ_{j_1\cdots j_{r-1}\check{j}_r}^{(r)x_r}=\bigotimes_{x_r=1}^{4}Z_{Lj_1\cdots j_{r-1},x_r}^{(r-1)}. \label{eq:stabilizerr_from_logicalr-1}
\end{align}
The logical-qubit operators at level $r$ are given by
\begin{align}
    &X_{Lj_1\cdots j_r}^{(r)}=\bigotimes_{x_r=3-j_r}^{4-j_r}X_{Lj_1\cdots j_{r-1},x_r}^{(r-1)}, \label{eq:logicalxr_from_logicalxr-1}\\
    &Z_{Lj_1\cdots j_r}^{(r)}=\bigotimes_{x_r=j_r}^{j_r+1}Z_{Lj_1\cdots j_{r-1},x_r}^{(r-1)}
\end{align}
for $j_r\in[1,2]$.
We choose the gauge-qubit operators within the hyperplane at $x_n=4$ ($n\in[1,r]$) so that there is no intersection between the gauge- and logical-qubits operators.

Specifically, the operators of the subsystem $D_{4,4,4}$ code are shown in Fig.\ref{fig:cube} (c).
The code is defined on $4^3=64$ physical qubits.
There are 8 logical qubits, the operators of which are defined on $2\times2\times2$ physical qubits, 
38 stabilizer generators consisting of 18 operators that extend along the $x$ direction and are defined on $4\times2\times2$ physical qubits, 12 along the $y$ direction on $2\times4\times2$ physical qubits, and 8 along the $z$ direction on $2\times2\times4$ physical qubits.
In addition, there are 18 gauge qubits.

\subsection{Encoding and syndrome measurements of $D_{4^r}$}

For encoding, one needs to fix the gauge to specify the encoded states.
Specifically, we encode a physical state within the codespace of the original $D_{4^r}$ codes following \cite{doi:10.1126/sciadv.adp6388}.
Due to the absence of the gauge qubits in the original codes, there is one-to-one correspondence between the physical and encoded states, that is, we can fix the gauge degrees of freedom.

We measure check operators parallel to each direction in order from $x_1$ to $x_r$.
That is, we first measure the $2\cdot 4^{r-1}$ check operators parallel to $x_1$, then measure those to $x_2$ and so on.
After measuring all the directions, a single step of the syndrome measurement is completed.
For each lines in the hyperlattice, we attach two ancilla qubits for the syndrome measurements of $X$ and $Z$ check operators, respectively, as in Fig.~\ref{fig:422level1}. 
We need $2\cdot 4^{r-1}$ ancilla qubits to measure all the check operators parallel to each other.
Since we can reuse the ancilla qubits for measuring check operators in the other directions, the space overhead due to the syndrome measurements is $3/2$.

The syndrome outcome is the eigenvalues of the stabilizer generators and agrees with the products of the measurement outcome of the check operators.
Hence, we need to measure only the weight-four check operators in the syndrome measurements.

\subsection{Subsystem $D_{6^r}$ codes}

The construction of the subsystem $D_{6^r}$ codes from the original $D_{6^r}$ codes is done in a similar way to that of the subsystem $D_{4^r}$ codes.
Compared with the original $D_{6^r}$ codes, the syndrome-measurement weight is reduced from $6\cdot 2^{r-1}$ to 6.

The $D_{6^r}$ codes are defined on qubits placed on an $r$-dimensional hyperlattice of linear dimension 5.
There are $r\cdot 6^{r-1}$ lines of length 5.
We define weight-6 check operators as the product of $X$ and $Z$ on these lines, which form the gauge group.
The procedure of constructing the subsystem $D_{6^r}$ codes from this point onward is similar to that of the subsystem $D_{4^r}$ codes in Sec.~\ref{sec:levelr}.
The resulting code has $4^r$ logical qubits, $2(5^r-4^r)$ stabilizer generators, and $6^r+4^r-2\cdot 5^r$ gauge qubits.
Specifically, the operators of the $D_{6,6}$, subsystem $D_{6,6}$, and subsystem $D_{6,6,6}$ codes are shown schematically in Figs.~\ref{fig:cube} (d)--\ref{fig:cube} (f).
We assign gauge-qubits operators within the hyperplanes at $x_n=4\,(n\in[1,r])$.

\section{Decoders}
\label{sec:decoder}

In the following, we focus on the subsystem $D_{4^r}$ codes and evaluate their performance
up to $r=4$ assuming bitflip errors.
The other errors such as measurement errors or circuit-level noise are not considered and left for future work. 
We used Stim \cite{gidney2021stim} to simulate the syndrome outcome.

Our codes have a large number of gauge qubits, the number of which grows exponentially with respect to the concatenation level.
Compared with the original $D_{4^r}$ codes, the number of the stabilizer generators decreases by that of the gauge qubits, that is, less information are available for decoding.
In addition, the number of error degeneracy is $2^\text{\# of the gauge qubits}$ as large as the $D_{4^r}$ codes since any two errors that are equivalent modulo gauge-qubit operators are regarded as the same error.
This makes optimal or nearly optimal decoders, which enumerate the all error degeneracy, infeasible at least within reasonable calculation time.
For this reason, we build a NN decoder to empirically learn the decoding up to level 4.
In addition, we use the block MAP decoder, which is optimal, up to level 3.

For comparison, we evaluate the logical error rate using the BP-OSD \cite{Panteleev2021degeneratequantum}, which is the standard for the qLDPC codes.
We used BP-OSD package \cite{PhysRevResearch.2.043423,Roffe_LDPC_Python_tools_2022} with iteration 200 and OSD order 2.
\begin{figure}
    \includegraphics[width=0.46\textwidth]{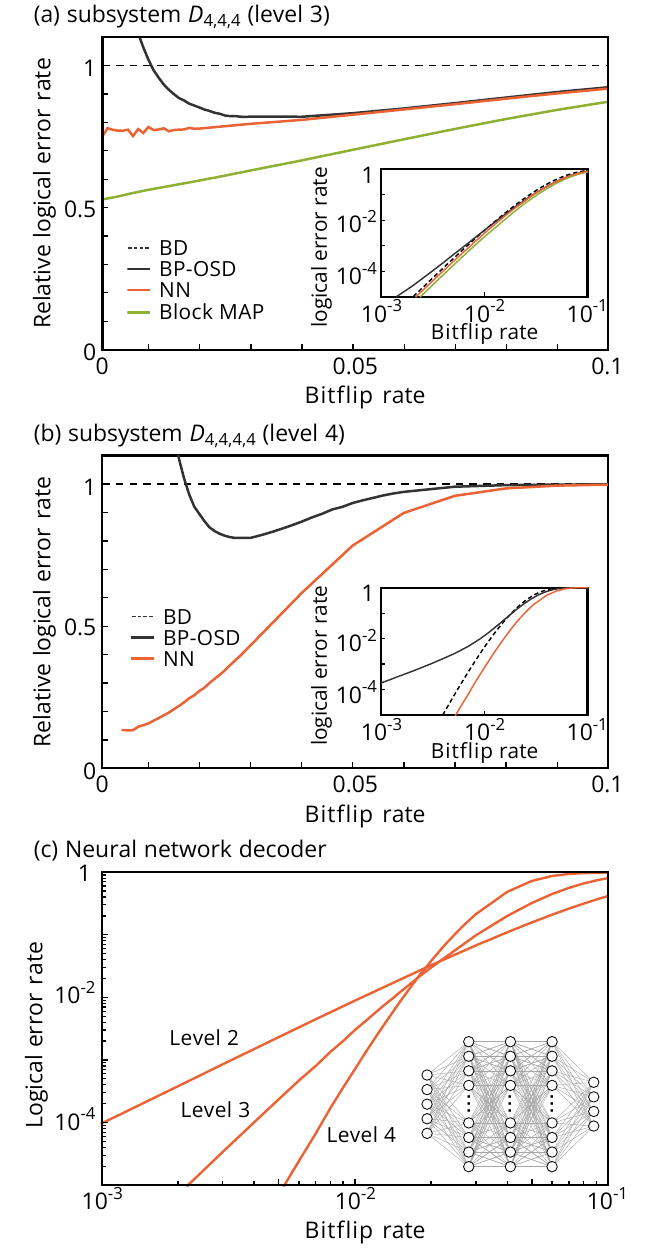}
    \caption{
        The performance of (a) the subsystem $D_{4,4,4}$ code and (b) the subsystem $D_{4,4,4,4}$ code, and (c) the subsystem $D_{4,4}$ to $D_{4,4,4,4}$ codes with the neural network (NN) decoder.
        In (a) and (b), the dashed lines are the theoretical performance of the bounded-distance (BD) decoder (\ref{eq:bdd_prob}) and solid lines are numerically evaluated by BP-OSD (black), the NN decoder (red), and the block MAP decoder (green), respectively.
        The vertical axis of (a) and (b) is the relative logical error rate compared with that of the BD decoder (\ref{eq:bdd_prob}), while the insets show the bare logical error rate in log scale.
        In (c), we plot the logical error rate estimated with the NN decoder from level 2 to 4. The inset shows the schematic picture of the NN at level 2.
        \label{fig:decoder}
    }
\end{figure}
All the results are compared with the theoretical performance of the BD decoder that successfully corrects up to $d/2-1$ errors, where $d=2^r$ is the code distance of the subsystem $D_{4^r}$ code.
The logical error rate of the BD decoder is given by
\begin{align}
    P^{(r)}(p)=1-\sum_{j=0}^{2^{r-1}-1}
    \dbinom{4^r}{j}p^j(1-p)^{4^r-j},
    \label{eq:bdd_prob}
\end{align}
where $p$ is the bitflip probability of each qubit.
When $p\ll 1$, $P^{(r)}(p)\propto p^{2^{r-1}}$.
The BP-OSD shows better performance than the BD decoder at high bitflip probability, while the logical error rate does not obey $p^{2^{r-1}}$ at lower $p$ [Fig.~\ref{fig:decoder} (a) and (b)].
The performance of the BP-OSD deteriorates as the concatenation level increases [Fig.~\ref{fig:decoder} (b)].

\subsection{Block MAP decoder}

Let us review the block MAP decoder, which deduces the most probable error $E\in\mathcal{P}_n$ by evaluating the maximum a posteriori (MAP) probability for a given syndrome outcome $\bm{s}$ \cite{PhysRevA.74.052333}, where $\mathcal{P}_n$ is the $n$-qubit Pauli group.
In evaluating the MAP probability, all the error operators that are equivalent modulo the gauge group needs to be identified.
In general, a bitflip error operator $E$ can be decomposed as
\begin{align}
    E=L(E)T(E)G(E),
    \label{eq:error_decomposition}
\end{align}
where $L(E)$ is a logical operator, $T(E)$ is a pure error operator, and $G(E)$ is an element of the gauge group $\mathcal{G}$.
Since Pauli operators $X$ and $Z$ are completely decoupled in our codes, all the components $L(E)$, $T(E)$, and $G(E)$ are composed solely of $X$ operators. 
The pure error $T(E)$ has a one-to-one correspondence with the syndrome outcome $\bm{s}$.
Specifically, $T(E)$ is a product of pure error operators $\{TX^{(r)x_n}_{j_1\cdots \check{j}_n\cdots j_r}\}$, each of which flips the eigenvalue of the stabilizer generator with the same indices following $\{SZ^{(r)x_n}_{j_1\cdots \check{j}_n\cdots j_r},TX^{(r)x_n}_{j_1\cdots \check{j}_n\cdots j_r}\}=0$, but has no other effect.
Here, $\{SZ^{(r)x_n}_{j_1\cdots \check{j}_n\cdots j_r}\}$ are the level-$r$ stabilizer generators in Eqs.~(\ref{eq:stabilizerr_from_stabilizerr-1}) and (\ref{eq:stabilizerr_from_logicalr-1}), the eigenvalues of which are the syndrome outcome $\bm{s}^{(r)}$.
The logical operator $L(E)$ is specified by a set of indices $\bm{l}^{(r)}$ by
\begin{align}
    L(E)
    =
    \prod_{j_1\cdots j_r}
    \left(X_{L j_1\cdots j_{r}}^{(r)}\right)^{l_{j_1\cdots j_{r}}^{(r)}}.
    \label{eq:logical_index}
\end{align}
As a result, the problem of finding the most probable error operator $E$ reduces to finding the maximum conditional probability $P(\bm{l}^{(r)}|\bm{s}^{(r)})$ of $\bm{l}^{(r)}$ from a given syndrome outcome $\bm{s}^{(r)}$.
Since $P(\bm{l}^{(r)}|\bm{s}^{(r)})=P(\bm{l}^{(r)},\bm{s}^{(r)})/\sum_{\bm{l}^{(r)}}P(\bm{l}^{(r)},\bm{s}^{(r)})$, the maximum conditional probability is equivalent to the maximum joint probability.
As the probability is evaluated exhaustively, the performance of the block MAP decoder must be optimal \cite{PhysRevA.74.052333}.

Here, we decompose the syndrome outcome $\bm{s}^{(r)}$ into $\bm{s}^{(r)x_1\cdots x_{r-1}}$ and $\bm{s}^{(r)x_r}$, which correspond to Eqs.~(\ref{eq:stabilizerr_from_stabilizerr-1}) and (\ref{eq:stabilizerr_from_logicalr-1}), respectively.
As we saw in Sec.~\ref{sec:levelr}, the level-$r$ operators are constructed from the level-$(r-1)$ ones. 
Let the indices of the level-$(r-1)$ codes be denoted by $\bm{l}^{(r-1)}$ and $\bm{s}^{(r-1)}$.
From Eqs.~(\ref{eq:stabilizerr_from_logicalr-1}) and (\ref{eq:logicalxr_from_logicalxr-1}), $\bm{s}^{(r)x_r}$ and $\bm{l}^{(r)}$ are determined by $\bm{l}^{(r-1)}$,
while $\bm{s}^{(r)x_1\cdots x_{r-1}}$ is by $\bm{s}^{(r-1)}$.
This reduces to a recursive relation of the joint probability given by \cite{PhysRevA.74.052333}
\begin{align}
    &P(\bm{l}^{(r)},\bm{s}^{(r)})
    =
    \sum_{\bm{l}^{(r-1)}}
    \delta(\bm{l}^{(r)},\bm{l}^{(r-1)})\delta(\bm{s}^{(r)x_r},\bm{l}^{(r-1)}) \notag\\
    &\times 
    \sum_{\bm{s}^{(r-1)}}\delta(\bm{s}^{(r)x_1\cdots x_{r-1}},\bm{s}^{(r-1)})
    P(\bm{l}^{(r-1)},\bm{s}^{(r-1)}),
    \label{eq:blockmap_poulin}
\end{align}
where $\delta$ denotes the indicator function, which is 1 when two sets of indices are consistent and 0 otherwise.

We numerically evaluate the joint probability and decode the subsystem $D_{4^r}$ codes up to $r=3$.
The logical error rate is proportional to $p^{2^{r-1}}$ and is smaller than the other decoders as it should be [Fig.~\ref{fig:decoder} (a)].
Decoding the subsystem $D_{4,4,4,4}$ code with the block MAP decoder was not available within a reasonable calculation time.

\subsection{Neural-network decoder}

For learning the decoding of the subsystem $D_{4^r}$ codes, we use fully-connected deep neural networks (NNs) \cite{Varsamopoulos_2018,Baireuther2018machinelearning,Chamberland_2018,PhysRevA.99.052351,8880492}.
The NNs learn the relation between the syndrome outcome and the indices of the most probable set of errors.
We assign the syndrome outcome $\bm{s}^{(r)}$ as the input and the indices $\bm{l}^{(r)}$ for the logical operator $L(E)$ in Eq.~(\ref{eq:logical_index}) as the output.
This class of decoders are called as high-level decoders in the sense that the output is not errors of the individual physical qubits but those of the logical qubits \cite{Varsamopoulos_2018}.
We train the NNs to minimize the binary cross entropy between the predicted $\bm{l}^{(r)}$, which takes real values between 0 and 1, and their groundtruth, which are a set of either 0 or 1.
As we assume bitflip errors, there are $3^r-2^r$ input nodes and $2^r$ output nodes at level $r$.

The NNs are trained using sets of the syndrome outcome and logical indices simulated by a circuit with a specific bitflip probability $p$.
The same NNs are used for decoding circuits with general $p$.
Specifically, we consider a fully-connected NN with 3 hidden layers with $16\times 3^r$ nodes.
The number of hidden layers and nodes are determined so that the performance is well converged at level $r=2,3$ and is the best at level $r=4$
\footnote{Within the number of samples used for training, NNs with more nodes and layers do not necessarily perform better and sometimes perform worse at level 4.}.
We trained the NNs by $2^{22}$ samples simulated with $p=0.04$ at level 2,
$2^{23}$ samples with $p=0.04$ at level 3, and
$2^{29}$ samples with $p=0.03$ at level 4.

The logical error rate of the NN decoder is better than the BD decoder, which indicates that the NN successfully corrects more errors than the bounded distance.
In addition, the NN decoder performs better than the BP-OSD across the entire range of $p$.
However, considering the performance difference between the block MAP and NN decoders at level 3 [Fig.~\ref{fig:decoder} (a)], the NN decoder may not fully utilize the potential of the subsystem $D_{4^r}$ codes, which we leave for future work.
The threshold for the bitflip error is estimated about 2\%.

\section{Conclusions}
\label{sec:conclusion}

In this work, we have proposed the subsystem MHC codes as subsystem codes of the MHC codes.
In the case of the $[\![4,2,2]\!]$ subsystem MHC codes, the weight of the check operators is reduced to four regardless of the concatenation level, and hence the problem of an exponentially large operator weight common to concatenated codes is resolved.
In addition, the net encoding rate including ancilla qubits at level $r$ is $1/(3\cdot 2^{r-1})$, which inherits high encoding rate from the original MHC codes.
The subsystem MHC codes proposed here can be extended to arbitrary MHC codes including those with the $[\![6,4,2]\!]$ code \cite{doi:10.1126/sciadv.adp6388}.

Due to the presence of exponentially growing number of gauge qubits, the numerical cost of exhaustive searches, such as the brute-force or block MAP decoding, grows far faster than the original $[\![4,2,2]\!]$ MHC codes as the concatenation level increases.
Thus, we have devd a fully-connected deep NN decoder to reduce the numerical cost in decoding.
The performance of the NN decoder exceeds that of the BD decoder and the BP-OSD.
Compared with the original $[\![4,2,2]\!]$ MHC codes, the estimated threshold for the bitflip error is reduced from 7\% to 2\% (see Appendix \ref{app:original}).
This reduction could be due to reduced information from the syndrome measurements, and also due to the decoder that is not sufficiently optimized.
Indeed, the performance of the NN decoder has not reached that of the block MAP (optimal) decoder.
Closing this gap would be a potential future work.

Compared with the teleportation-based error correction proposed in \cite{doi:10.1126/sciadv.adp6388}, %which uses twice as large number of ancillae as that of the data qubits, 
the advantages of the proposed syndrome extraction by ancillae measurements are that the number of ancillae is reduced by a factor of 1/4 (1/6) for the subsystem $D_{4^r}$ ($D_{6^r}$) codes,
and that this method is free from preparing logical Bell states.
However, the syndrome extraction with ancillae in general requires repeated measurements, which would make the time overhead larger than the teleportation-based one.
In addition, entangling data qubits with ancillae could lead to error spreading, which is not considered in this work.
We will discuss the circuit-level noise separately in the forthcoming paper.

\begin{acknowledgements}
    This work was supported by JST Moonshot R\&D Grant Number JPMJMS2061.
\end{acknowledgements}

\appendix

\section{$D_{4^r}$ codes}
\label{app:original}

In this section, we show the performance of the original $D_{4^r}$ codes.

The original and subsystem $D_{4^r}$ codes share the same logical-qubit operators,
while the stabilizer generators are slightly different.
Unlike the subsystem $D_{4^r}$ codes, the original one inherits the lower-level stabilizer generators after concatenation.
In total, there are $4^r-2^r$ stabilizer generators, which constitute a $[\![4^r,2^r,2^r]\!]$ stabilizer code.

\begin{figure}
    \includegraphics[width=0.47\textwidth]{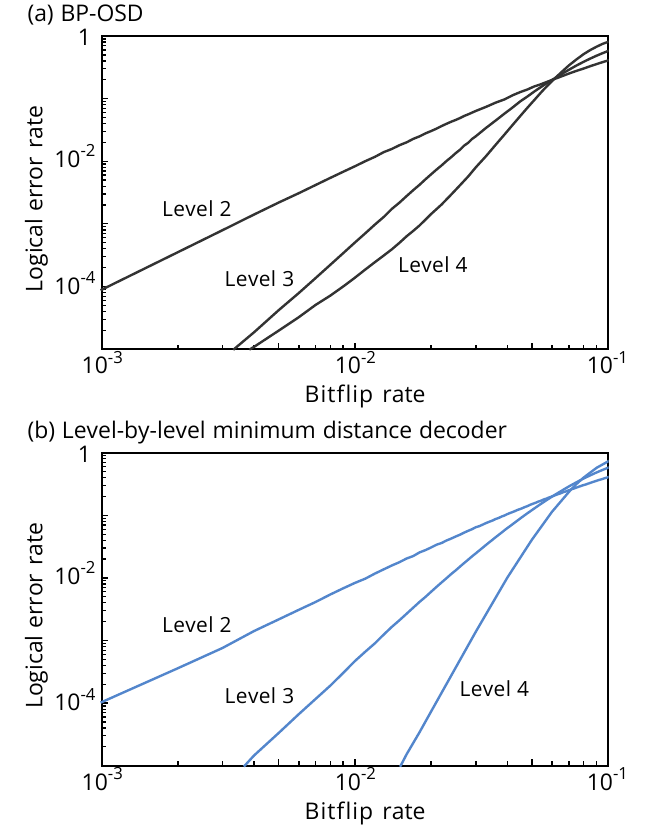}
    \caption{
        The performance of the $D_{4^r}$ codes by (a) the BP-OSD and (b) the level-by-level minimum distance decoder \cite{doi:10.1126/sciadv.adp6388}.
        Numerically evaluated logical error rates are plotted from level 2 to 4.
        \label{fig:decoder_stabilizer}
    }
\end{figure}

We numerically evaluate the performance of the $D_{4^r}$ codes assuming bitflip error (Fig.~\ref{fig:decoder_stabilizer}).
We employ the BP-OSD and the level-by-level minimum distance (MD) decoder \cite{doi:10.1126/sciadv.adp6388}.
The threshold for the bitflip error estimated with the MD decoder is about 7\%.

\bibliography{main.bbl}

\end{document}